\documentclass[%
 preprint, %linenumbers,
superscriptaddress,
 amsmath,amssymb,
 aps, physrev,
]{revtex4-2}

\usepackage{graphicx}% Include figure files
\usepackage{dcolumn}% Align table columns on decimal point
\usepackage{bm}% bold math
\usepackage[version=4]{mhchem}
\usepackage{chemfig}

\begin{document}

\preprint{APS/123-QED}

\title{Discovery of the Type-II Superconductor Ta$_4$Rh$_2$C$_{1-\delta}$ with a High Upper Critical Field}

\author{KeYuan Ma}
\affiliation{Max Planck Institute for Chemical Physics of Solids, 01187 Dresden, Germany}

\author{Sara López-Paz}
\affiliation {Department of Quantum Matter Physics, University of Geneva, CH-1211 Geneva, Switzerland}

\author{Karolina Gornicka}
\affiliation {Faculty of Applied Physics and Mathematics and Advanced Materials Centre, Gdansk University of Technology, Gdansk 80-233, Poland}
\affiliation{Department of Quantum Matter Physics, University of Geneva, CH-1211 Geneva, Switzerland}

\author{Harald O. Jeschke}
\affiliation {Research Institute for Interdisciplinary Science, Okayama University, Okayama 700-8530, Japan}

\author{Tomasz Klimczuk}
\affiliation {Faculty of Applied Physics and Mathematics and Advanced Materials Centre, Gdansk University of Technology, Gdansk 80-233, Poland}

\author{Fabian O. von Rohr}
%\email{fabian.vonrohr@unige.ch}
\affiliation{Department of Quantum Matter Physics, University of Geneva, CH-1211 Geneva, Switzerland}

\begin{abstract}

 We report on the discovery of superconductivity in the previously unknown compound Ta$_4$Rh$_2$C$_{1-\delta}$. Ta$_4$Rh$_2$C$_{1-\delta}$ crystallizes in the $\eta$-carbide structure type, in the cubic space group $Fd\bar{3}m$ (No.227) with a unit cell parameter of $a = $ 11.7947 \AA. Temperature-dependent magnetic susceptibility, resistivity, and specific heat capacity measurements reveal that Ta$_4$Rh$_2$C$_{1-\delta}$ is a type-II bulk superconductor with a critical temperature of $T_{\rm c}$ = 6.4 K, and a normalized specific heat jump $\Delta C/\gamma T_{\rm c}$ = 1.56. Notably, we find Ta$_4$Rh$_2$C$_{1-\delta}$ has a high upper critical field of $\mu_0 H_{\rm c2}{\rm (0)}$ = 17.4 T, which is exceeding the BCS weak coupling Pauli limit of $\mu_0 H_{\rm Pauli}$ = 11.9 T. 

\end{abstract}

\maketitle

\section{Introduction}

The discovery of new superconductors with enhanced properties for diverse applications remains a significant challenge in condensed matter physics \cite{larbalestier2011superconductivity,von2023chemical,simon1997superconductivity,santoro1988crystal}. A crucial property for these applications is the upper critical field $\mu_0 H_{\rm c2}(0)$, which is essential for technological applications \cite{hahn201945}. When an external magnetic field is applied to a superconductor, Cooper pairs may break due to two effects: the orbital-limiting effect, which induces a momentum leading to a supercurrent that exceeds the superconducting gap, and the Pauli paramagnetic effect (Zeeman effect), where the Zeeman energy surpasses the superconducting condensation energy \cite{Tinkham2004}. Near the critical temperature $T_{\rm c}$, the orbital-limiting effect dominates, while the Pauli paramagnetic effect is more significant at lower temperatures. In BCS theory, the maximum $\mu_0 H_{\rm c2}(0)$ is limited by the Pauli paramagnetic effect, known as the Pauli paramagnetic limit $\mu_0 H_{\rm Pauli}$, given as $\mu_0 H_{\rm Pauli} \approx 1.86{\rm [T/K]} \cdot T_{\rm c}$ \cite{Tinkham2004}. Several superconductors with the $\eta$-carbide type crystal structure have recently been found to violate the Pauli paramagnetic limit, exhibiting very high upper critical fields \cite{ma2021superconductivity,ma2021group,watanabe2023observation}. 

$\eta$-carbide type compounds crystallize in the cubic space group $Fd\bar{3}m$, and commonly form with compositions of $A_4B_2X$ and $A_3B_3X$ where \textit{A} and \textit{B} stand for transition metals, and \textit{X} for carbon, nitrogen, or oxygen \cite{kuo1953formation,ku1984new,gupta2009structural,mackay1994new}. $\eta$-carbide type compounds consist of more than 100 known members with combinations of different technologically useful properties such as high hardness, high thermal stability, rich variety of magnetic states, exotic electronic properties, and catalytic properties \cite{waki2010itinerant,prior2004superparamagnetism,cui2020high}. One of the most striking characteristics of $\eta$-carbide compounds is that they exist over wide ranges of chemical compositions and allow for a high degree of atomic substitutions \cite{taylor1952new}. In this structure type, tuning of the chemical composition allows for modifying and controlling of the physical properties in a wide range. Hence, the flexibility and tunability of the $\eta$-carbide structure provide numerous opportunities to achieve new quantum materials with intriguing physical properties. 

Among the systematically investigated $\eta$-carbide type superconductors, Ti$_4$Co$_2$O, Ti$_4$Ir$_2$O, Nb$_4$Rh$_2$C$_{1-\delta}$, and Zr$_4$Pd$_2$O were found to have $\mu_0 H_{\rm c2}{\rm (0)}$ larger than the weak coupling Pauli limit, where $\mu_0 H_{\rm Pauli} \approx 1.86{\rm [T/K]} \cdot T_{\rm c}$ \cite{ma2021superconductivity,ma2021group,watanabe2023observation,ruan2022superconductivity}. These isostructural superconductors share many electronic property features; therefore, it is likely they also share a common origin for the unusually high upper critical fields. Recently, in the high-field region of Ti$_4$Ir$_2$O signatures for a Fulde-Ferrell-Larkin-Ovchinnikov state have been observed, and $\mu$SR measurements have revealed a small superfluid density in the superconducting state of Ti$_4$Ir$_2$O \cite{das2024ti}. Both observations point towards unconventional superconductivity in this family of materials. Therefore, the $\eta$-carbide family of compounds has become a fertile ground for the discovery of novel superconducting materials.

Here, we report on the discovery of superconductivity in the previously unreported $\eta$-carbide compound Ta$_4$Rh$_2$C$_{1-\delta}$. We find Ta$_4$Rh$_2$C$_{1-\delta}$ to crystallize in the $\eta$-carbide structure type with a unit cell parameter of $a = $ 11.7947 \AA. Furthermore, we show that this compound is a type-II bulk superconductor with a critical temperature of $T_{\rm c} $ = 6.4 K, and a specific heat jump $\Delta C/\gamma T_{\rm c}$ of 1.56. Moreover, we find that Ta$_4$Rh$_2$C$_{1-\delta}$ -- like some other $\eta$-carbide superconductors -- has a very high upper critical field of $\mu_0 H_{\rm c2}(0)$ of 17.4 T, which exceeds the weak coupling Pauli paramagnetic limit $\mu_0 H_{\rm Pauli}$ of 11.9 T.

\section{EXPERIMENTAL DETAILS }

\textbf{Synthesis:}
Polycrystalline Ta$_4$Rh$_2$C$_{1-\delta}$ was synthesized from nearly stoichiometric amounts of the elements using tantalum powder (99.99 \%, Alfa Aesar), rhodium powder (99.95 \%, Strem Chemicals), and carbon rod (99.999 \%, Sigma-Aldrich). A total mass of 150 mg of starting material was used. The reactants were thoroughly mixed in an agate mortar and pressed into a pellet. The pellet was first melted in an arc furnace in a purified argon atmosphere on a water-cooled copper plate. The sample was flipped over and molten five times to ensure an optimal homogeneity. After arc-melting, only a small mass loss of approximately 1 \% was observed. The very hard solidified melt ingot was crushed into small particles in a tungsten carbide mortar and ground to fine powders in an agate mortar and pressed into a pellet. The pellet was wrapped with thin Ta foil, sealed in a quartz tube under a 1/3 partial argon, and annealed in a furnace for 4 days at 1200 $^\circ$C. After reaction, the quartz tube was cooled down to room temperature by quenching in water.
    
\textbf{Structure and Composition:}
 The crystal structure and phase purity of the sample were checked using powder X-ray diffraction (PXRD) measurements on a Rigaku SmartLab diffractometer with Cu K$_{\alpha}$ radiation in Bragg-Brentano reflection geometry. The PXRD patterns were collected in the 2$\Theta$ range of 5 -120$^{\circ}$ with a scan rate of 0.25$^{\circ}$/min. Rietveld refinements were performed using the FULLPROF  program package \cite{rodriguez2001recent}. The chemical composition of the polycrystalline samples were examined under a scanning electron microscope (SEM) (JEOL JSM-IT800 operated at 15 keV) equipped with an energy-dispersive X-ray (EDX) spectrometer. 

\textbf{Physical Property Measurements:}
Temperature- and field-dependent magnetization measurements were performed on a Quantum Design magnetic properties measurement system (MPMS3) with a 7 T magnet equipped with a vibrating sample magnetometry (VSM) option. The measured pellet was placed parallel to the external magnetic field to minimize the demagnetization effects in the superconducting state. The electrical resistivity and specific heat capacity measurements were conducted in a Quantum Design physical property measurement system (PPMS) with a 9 T magnet. For the resistivity measurements, the four-probe technique was employed with gold wires connected to the sample with silver paint. Specific heat measurements were performed with the Quantum Design heat-capacity option, using a relaxation technique.  

\textbf{Electronic Structure Calculations:} We performed density functional theory (DFT) calculations based on the full potential local orbital (FPLO) basis set \cite{koepernik1999full} to understand the electronic structure of Ta$_4$Rh$_2$C$_{1-\delta}$ . Due to heavy elements Ta and to some extent Rh, spin-orbit coupling is expected to be strong in Ta$_4$Rh$_2$C$_{1-\delta}$ , and we use fully relativistic calculations with the generalized gradient approximation exchange correlation functional \cite{perdew1996generalized} to account for the spin-orbit coupling effects in the electronic structure. We converge the calculations on $16\times 16\times 16$ $k$ meshes.

\section{RESULTS and DISCUSSION}

\begin{figure}[h!]
\includegraphics[width=0.7\textwidth]{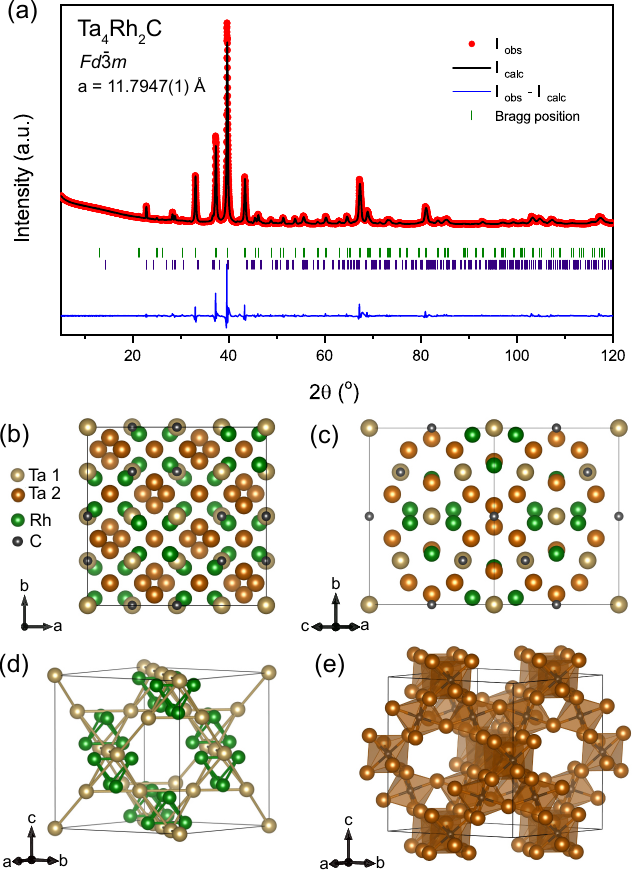}
\caption{ (a) Rietveld refinements of the room temperature PXRD pattern of  Ta$_4$Rh$_2$C$_{1-\delta}$. The plots are represented as follows: observed (red colored dots), calculated (black colored line), and difference (blue colored line) intensities. The Bragg positions of main Ta$_4$Rh$_2$C$_{1-\delta}$ phase (96.5(4) \%), and Ta$_2$O$_5$ impurity phase (3.5(1) \%) are indicated with green, and purple colored vertical ticks, respectively. (b)-(e) Schematic representation of different orientations for the refined crystal structure from PXRD of Ta$_4$Rh$_2$C$_{1-\delta}$}
\label{fig1}
\end{figure}

\section{Synthesis}

To the best of our knowledge, no compound in the ternary Ta-Rh-C phase space has been previously reported. Here, we report the $\eta$-carbide compound Ta$_4$Rh$_2$C$_{1-\delta}$ in this system. We have obtained polycrystalline Ta$_4$Rh$_2$C$_{1-\delta}$ as a silver colored pellet. Achieving a phase-pure Ta$_4$Rh$_2$C$_{1-\delta}$ sample proved to be challenging. Direct high-temperature reactions of mechanically mixed elements pressed into a pellet produced low-quality samples with multiple phases. This is likely due to the difficulty in homogeneously mixing the grains of the reactants. We found that arc-melting the reactants was crucial, even though the $\eta$-carbide phase was not present in the ingot immediately after arc-melting, but seems to allow for optimal mixing of the elements (see supporting information S-Fig.1) \cite{supp}. In preparative chemistry, the synthesis of phase pure ternary $\eta$-carbides is generally known to be challenging \cite{ku1984new,ku1985effect}. In a series of synthesis experiments, the highest purity final samples were obtained with a starting ratio of Ta:Rh:C as 3.85:2.15:0.85. Deviation from this compositional ratio or from the annealing temperature (1200 $^\circ$C) resulted in the formation of significant impurity phases.

\subsection{Crystal Structure}

We find Ta$_4$Rh$_2$C$_{1-\delta}$ to crystallize in the $\eta$-carbide type structure, with the cubic space group $Fd\bar{3}m$ (no.227) with the cell parameter of $a = $ 11.7947(1) \AA. This structure can be rationalized as related compounds, e.g. Nb$_4$Rh$_2$C$_{1-\delta}$ ($a = $ 11.8527(2) \AA) crystallize also in it. Here, the powder X-ray diffraction technique was employed to identify the phase purity and cell parameters of the obtained samples. 

The powder X-ray diffraction (PXRD) pattern and the corresponding Rietveld refinement of the obtained Ta$_4$Rh$_2$C$_{1-\delta}$ sample are presented in Figure \ref{fig1}(a). Energy-dispersive X-ray spectroscopy (EDS) analysis reveals a Ta:Rh ratio of 1.9(5):1 for the sample, which is close to the ideally stoichiometric value of 2:1. Reliable quantification of the carbon content by EDS is not possible and is challenging by X-ray diffraction as well \cite{supp}. Assuming only a negligible carbon loss during the arc-melting process, the carbon content in Ta$_4$Rh$_2$C$_{1-\delta}$ should be close to 0.85, i.e. the nominal composition, for the amount of carbon used for the purest obtained sample. Rietveld refinement analysis determined that the main phase, Ta$_4$Rh$_2$C$_{1-\delta}$, constitutes 96.5 \% of the sample, with a minor impurity phase of Ta$_2$O$_5$ at 3.5 \%. Notably, the formation of TaC as an impurity phase was not observed for these synthesis conditions. Details on the Rietveld refinements of Ta$_4$Rh$_2$C are summarized in Table~\ref{table:Table1}, assuming a model $\eta$-carbide structure type, with the cubic space group $Fd\bar{3}m$ and Ta$_4$Rh$_2$C stoichiometry.

\begin{table} [h!]
\caption{Atomic and cell parameters of Ta$_4$Rh$_2$C$_{1-\delta}$ obtained from Rietveld refinement of the room temperature PXRD data.}
		\begin{tabular}{| c | c | c | c | c | c | c |}
\hline
Atom  &  Site & $x$ & $y$  & $z$ & $B_{iso}$ & $Occ.$ \\ \hline
 Ta1 & $16c$ & 0 & 0 & 0 & 1.13(1) & 1 \\
 Rh & $32e$ & 0.21185(7) & 0.21185(7) & 0.21185(7) & 0.77(3) & 1 \\
 Ta2 & $48f$  & 0.44067(7) & 0.125 & 0.125 & 1.13(1) & 1 \\
 C & $16d$ & 0.5 & 0.5 & 0.5 & 1.1* & 1 \\ \hline
\multicolumn{7}{|c|}{$Fd\bar{3}m$ (no.227);  $a$ = 11.7947(1) \AA } \\
\multicolumn{7}{|c|}{$R_p$ (\%) = 5.97:  $R_{wp}$ (\%) = 9.56;  $R_{Bragg}$ (\%)  = 5.43} \\ \hline
\multicolumn{7}{|c|}{\small Note : Here, *$B_{iso}$ is fixed to the refined overall value} \\ \hline
\end{tabular}
\label{table:Table1}
\end{table} 

In Figure \ref{fig1}(b)-(e), we show the crystal structure of Ta$_4$Rh$_2$C$_{1-\delta}$ in an ideal chemical stoichiometric general formula of Ta$_4$Rh$_2$C. In this $\eta$-carbide structure, Ta atoms occupy the $16c$ and the $48f$ Wyckoff positions, Rh atoms occupy the $32e$ Wyckoff positions, and C atoms occupy the $16d$ Wyckoff positions. Even though there are only 4 Wyckoff positions required to describe the crystal structure, the unit contains nevertheless results in a total of 112 atoms and a formula of Ta$_{64}$Rh$_{32}$C$_{16}$ for one unit cell. In Figure \ref{fig1}(b)\&(c) the unit cell with all atoms are shown in two orientations. In \ref{fig1}(d), the connectivity of the Ta1 and Rh atoms are shown: the Ta1 atoms form a network of tetrahedra resulting in a stella quadrangla structure, while the Rh atoms arrange in isolated tetrahedra. The Ta 2 atoms form a network of octahedra in which every second one is slightly distorted, as shown in figure \ref{fig1}(e), with the C atoms filling the octahedral voids.

\subsection{Physical Properties}

To understand the physical properties of $\eta$-carbide compound Ta$_4$Rh$_2$C$_{1-\delta}$, we performed temperature dependent magnetic susceptibility, resistivity, and specific heat capacity measurements. 

\begin{figure}[h!]
\centering
\includegraphics[width=0.5\linewidth]{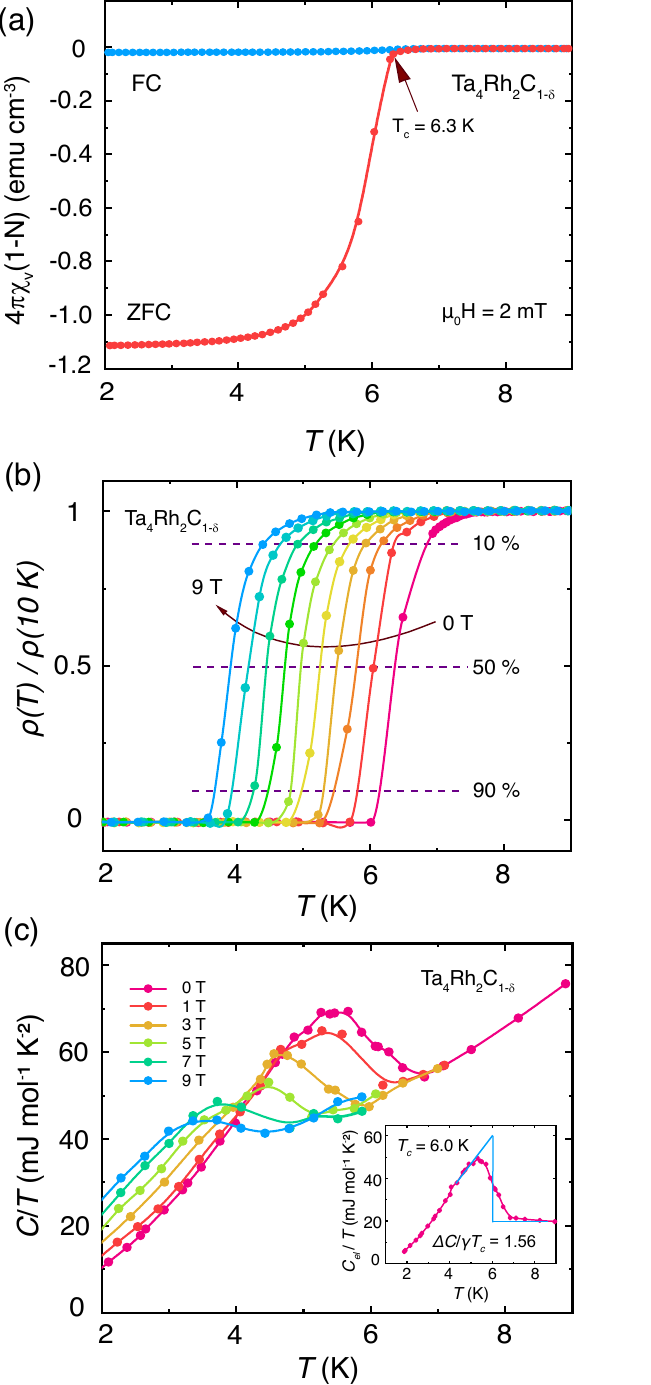}
\caption{Superconducting properties of Ta$_4$Rh$_2$C$_{1-\delta}$. (a) Zero-field cooled (ZFC) and field cooled (FC) temperature-dependent magnetic susceptibility in a temperature range from $T$ = 2 K to 9 K. (b) Field-dependent resistivity in the vicinity of the superconducting transition in fields between $\mu_0 H =$ 0 T and 9 T. (c) Field dependent specific heat in fields between $\mu_0 H =$ 0 T and 9 T. Inset: Entropy-conserving construction of the zero field measurement.}
\label{fig2}
\end{figure}

In Figure \ref{fig2}(a), we observe a superconducting transition at a critical temperature of $T_{\rm c}$  = 6.3 K in the temperature dependence of the magnetic susceptibility in zero-field cooled (ZFC) and field-cooled (FC) modes under an external field of $\mu_0 H =$ 2 mT, respectively. The difference between the FC and ZFC measurements in the superconducting state are prototypical for a type-II superconductor. When dealing with the magnetic susceptibility data, a demagnetization factor N was estimated using the relationship -b = 1/[4$\pi$ (1-N)] \cite{carnicom2018tarh2b2}. Here, we obtain a value of N = 0.53 for our sample by fitting the field-dependent measurements of the magnetization to a line ($M$ = b$H$ + a) in the low-field region \cite{carnicom2018tarh2b2}.  

In the normal state, Ta$_4$Rh$_2$C$_{1-\delta}$ shows a Pauli paramagnetic behavior, as confirmed by magnetization measurement between 10 K to 300 K in an external field of $\mu_0 H =$ 1 T (see supporting information S-Fig.3) \cite{supp}. To estimate the lower critical field $H_{\rm c1}$, we performed a series of field-dependent measurements of the magnetization in low fields below the critical temperature $T_{\rm c}$, as shown in S-Fig 4.(a) of the supporting information \cite{supp}. Here, we used the magnetic-field point where the $M$($H$) curve first deviates from linearity as the measure for $H_{\rm c1}$ \cite{naito1990temperature}. With this approximation, the obtained $H_{\rm c1}$ values are fitted using the semi-empirical formula:
\begin{equation}
H_{c1}(T) = H_{c1}(0) [1-(T/T_c)^2]. 
\end{equation}

The lower critical field at $T =$ 0 K is determined to be $\mu_0 H_{\rm c1}(0) $= 9.4 mT as shown in the S-Fig 4 (b) (see supporting information) \cite{supp}. After taking the demagnetization factor N = 0.53 into account, the lower critical ﬁeld is corrected to be $\mu_0 H_{\rm c1}^{*}(0) $= 20 mT.

We find the resistivity of Ta$_4$Rh$_2$C$_{1-\delta}$ to decrease with decreasing temperature, showing a metallic behavior. The temperature dependent electrical resistivity measurement of the polycrystalline Ta$_4$Rh$_2$C$_{1-\delta}$ sample from 300 to 1.8 K is shown in the supporting information (see S-Fig. 5) \cite{supp}. At a critical temperature $T_{\rm c, onset}$ of 7.2 K, Ta$_4$Rh$_2$C$_{1-\delta}$ starts to undergo a transition to a superconducting state and the resistivity completely drops to zero at 6.0 K. Here, the residual resistivity ratio (RRR) value of the annealed polycrystalline sample is defined as $\rho$(300 K)/$\rho$(10 K) $\approx$ 1.22, corresponding to a poor metal behavior. This small RRR value may arise from the polycrystalline nature of the sample. In Figure \ref{fig2}(b), we show the temperature- and field-dependent resistivity $\rho$($T$,$H$) in a temperature range from $T$ = 2 to 9 K and magnetic fields between $\mu_0 H =$ 0 T and 9 T. We show that the resistivity drops to zero at the transition to the superconducting state for all applied fields. In zero field, we determine the critical temperature to be $T_{\rm c}$ = 6.4 K with a 50 \% criterion. As expected, the critical temperature decreases steadily as the applied magnetic field increases. However, the critical temperature is only suppressed to $T_{\rm c} =$ 3.9 K in the maximal applied field of 9~T, which is already evidence of the remarkably high upper critical field of this superconductor.   

In Figure \ref{fig2}(c), we present the temperature- and field-dependent specific heat $C$($T$,$H$) in the vicinity of the superconducting transitions, where the data are plotted as $C/T$ versus $T$ in magnetic fields between $\mu_0 H =$ 0 T and 9 T. The specific heat jumps corresponding to the superconducting transitions are well-pronounced in all measured fields, and shift to lower temperatures, which are in good agreement with the results from the resistivity measurements. At zero field, the specific heat jump temperature corresponding to the superconducting transition is determined to be $T_{\rm c}$ = 6.0 K based on an entropy-conserving construction, as shown in the inset of Figure \ref{fig2}(c). 

In the normal state -- close to the superconducting transition -- the specific heat can be fitted according to the expression: 

\begin{equation}
\frac{C(T)}{T} = \frac{C_{el} + C_{ph}}{T} = \gamma + \beta T^2
\end{equation}

where $\gamma$ is the Sommerfeld coefficient, corresponding to the electronic contribution to $C$($T$), and $\beta$$T^3$ is the phonon  contribution to the specific heat. Here, we obtain the $\gamma$ to be 20.9 mJ mol$^{-1}$ K$^{-2}$, and the  $\beta$ to be 0.707 mJ mol$^{-1}$ K$^{-4}$ for Ta$_4$Rh$_2$C$_{1-\delta}$ (see supporting information S-Fig.6) \cite{supp}. With above obtained $T_{\rm c}$ and $\gamma$ values, the normalized specific heat jump is found to be $\Delta C/\gamma T_{\rm c}$ = 1.56 in zero field, which is slightly larger than the weak-coupling BCS value of 1.43 and evidence for the bulk nature of the superconducting state in Ta$_4$Rh$_2$C$_{1-\delta}$.

We determined the Debye temperature to be $\Theta_D =$ 268 K, using the following relationship:  
\begin{equation}
\Theta_D = \left(\frac{12 \pi^4}{5 \beta} n R \right)^{\frac{1}{3}} 
\end{equation}

Here n = 7 is the number of atoms per formula unit, and R = 8.314 J mol$^{-1}$ K$^{-1}$ is the ideal gas constant.

The electron-phonon coupling constant $\lambda_{\rm ep}$ can be estimated from the Debye temperature, using the semi-empirical McMillan approximation \cite{mcmillan1968transition}: 
\begin{equation}
\lambda_{\rm ep} = \dfrac{1.04 + \mu^{*} \ {\rm ln}\big(\frac{\Theta_{\rm D}}{1.45 T_{\rm c}}\big)}{(1-0.62 \mu^{*}) {\rm ln}\big(\frac{\Theta_{\rm D}}{1.45 T_{\rm c}}\big)-1.04}.
\end{equation} 
Here, the Coulomb repulsion parameter $\mu^{*}$ is set to be 0.13 according to an empirical approximation that was widely used in superconductors with similar elements (e.g. NbRh$_2$B$_2$ and TaRh$_2$B$_2$) \cite{von2016effect, carnicom2018tarh2b2,gornicka2021nbir2b2,gui2022lair3ga2, zeng2022ta}. Based on these values, the $\lambda_{\rm ep} $ value for Ta$_4$Rh$_2$C$_{1-\delta}$ is calculated to be 0.71, which is smaller than the 0.83 for Nb$_4$Rh$_2$C$_{1-\delta}$.

The measured $\gamma$ value  corresponds to a density of states at the Fermi-level of $D(E_{\rm F})$ of 5.23 states eV$^{-1}$ per formula unit (f.u.) in Ta$_4$Rh$_2$C$_{1-\delta}$, when using the following relationship:

\begin{equation}
D(E_{\rm F}) = \dfrac{3 \gamma}{\pi^2 k_{\rm B}^2 (1+\lambda_{\rm ep})}.
\end{equation}

\subsection{High upper critical field in Ta$_4$Rh$_2$C$_{1-\delta}$}

\begin{figure}[h!]
\centering
\includegraphics[width=0.6\linewidth]{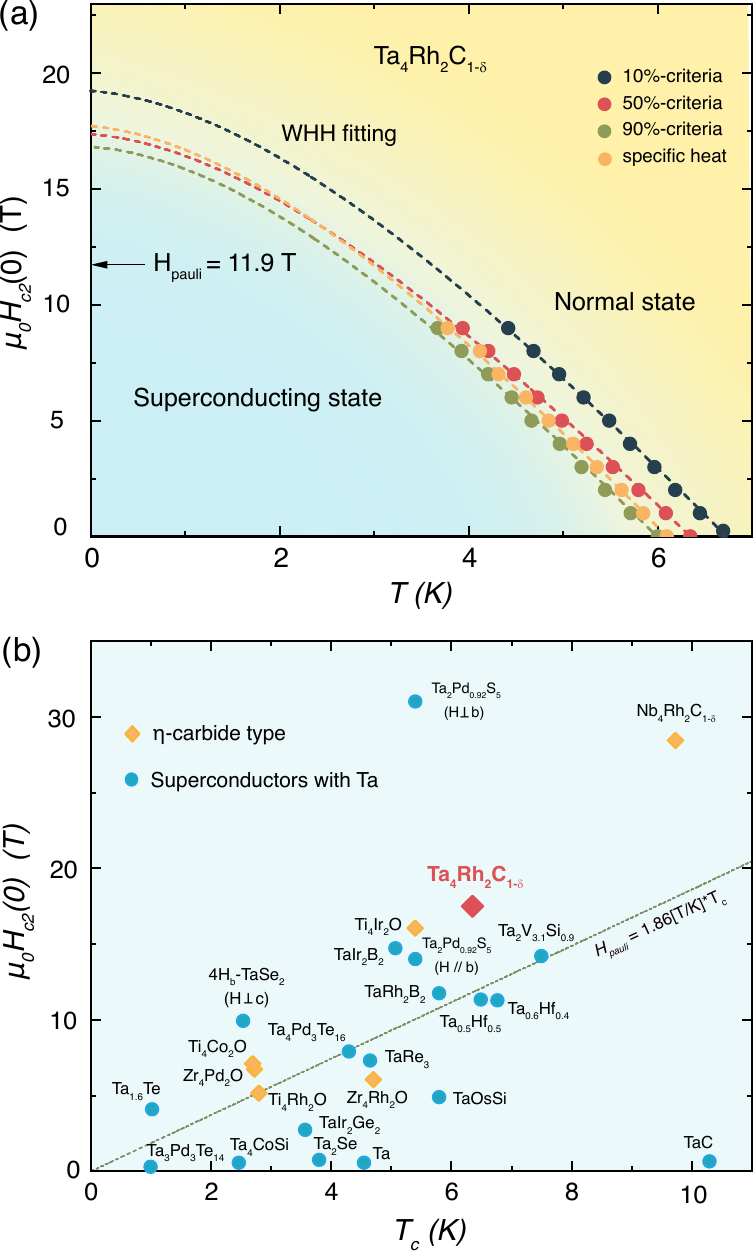}
\caption{(a) Upper critical field $\mu_0H_{\rm c2}$ of Ta$_4$Rh$_2$C$_{1-\delta}$. Data points from field and temperature dependent resistivity using the 10\%, 50\%, 90\%-criteria, and specific heat capacity measurements are shown. The data points were fitted using equation \ref{WHH}. The presented weak-coupling BCS Pauli limit $H_{\rm Pauli} \approx 1.86{\rm [T/K]} \times T_{\rm c}$ was calculated using $T_{\rm c}$ values from the 50\% criterion of resistivity. (b) Comparison of Ta$_4$Rh$_2$C$_{1-\delta}$ in $T_{\rm c}$ and $\mu_0 H_{\rm c2}(0)$ with previously reported $\eta$-carbide type superconductors and superconductors containing Ta.}
\label{fig3}
\end{figure}

In Figure \ref{fig3}(a), we present the field-dependence of the critical temperatures $T_{\rm c}$ determined from the resistivity measurements using the common 10 \%-, 50 \%-, and 90 \%-criteria , as well as the specific heat capacity as shown in Figures \ref{fig2}(b) and (c), (compare, e.g., references \cite{ma2021superconductivity,von2016effect,stolze2018sc}). When using the Ginzburg-Landau (GL) formalism, the zero-temperature upper-critical field $\mu_0 H_{\rm c2}$(0) for Ta$_4$Rh$_2$C$_{1-\delta}$ is determined to be 23.5 T, 20.7 T, 20.1 T, and 21.3 T for the 10 \%-, 50 \%-, and 90 \%-criteria of resistivity and the specific heat capacity, respectively (see supporting information S-Fig.7) \cite{supp}. Usually, a Ginzburg-Landau (GL) model fitting will give higher estimated upper-critical field values for the $\eta$-carbide structure type superconductors \cite{ma2021superconductivity,ma2021group,das2024ti}.

Here, we make a conservative estimation of the upper-critical field $\mu_0 H_{\rm c2}{\rm (0)}$ for Ta$_4$Rh$_2$C$_{1-\delta}$ using the Werthamer-Helfand-Hohenberg (WHH) formalism in the dirty limit according to \cite{helfand1966temperature, ma2021superconductivity,baumgartner2013effects}:

\begin{equation}
\label{WHH}
\mu_0 H_{c2} (T)= \frac{\mu_0 H_{c2} (0)}{0.693} h^\ast_{fit} (t).
\end{equation}
with $h^\ast_{fit}$ being
\begin{equation}
 h^\ast_{fit} (t)=(1-t)-C_{1}(1-t)^2-C_{2}(1-t)^4.
\end{equation}

where $t$ = $T/T_c$ ($T_c$ is the transition temperature at zero field), while $C_{1}$ = 0.153 and $C_{2}$ = 0.152 are two parameters \cite{baumgartner2013effects}.  The zero-temperature upper-critical field $\mu_0 H_{\rm c2}$(0) is determined to be 19.3 T, 17.4 T, 16.9 T, and 17.7 T for the 10 \%-, 50 \%-, and 90 \%-criteria of resistivity and the specific heat capacity, respectively. All of these values exceed the corresponding weak-coupling BCS Pauli paramagnetic limits of $\mu_0 H_{\rm Pauli} \approx 1.86{\rm [T/K]} \times T_{\rm c} = $ 12.5 T, 11.9 T, 11.1 T, and 11.2 T, respectively. 

In Figure \ref{fig3}(b), we present a comparison of Ta$_4$Rh$_2$C$_{1-\delta}$ with previously reported superconductors containing Ta in $T_{\rm c}$ and $\mu_0 H_{\rm c2}(0)$ evaluation \cite{carnicom2018tarh2b2,srivichitranond2017superconductivity,jiao2016superconductivity,xu2019two, shi2020incommensurate, zeng2022ta, shang2020superconductivity, klimczuk2023superconductivity, barker2018superconducting, gui2020superconductivity, gornicka2021nbir2b2, lu2014superconductivity, yan2023modulating, liu2023vanadium, terashima2024anomalous}. The critical temperature $T_{\rm c}$ of Ta$_4$Rh$_2$C$_{1-\delta}$ is higher than most of the reported Ta-based superconductors and its $\mu_0 H_{\rm c2}(0)$ value is higher than all the listed known superconductors except the highly anisotropic Ta$_2$Pd$_{0.92}$S$_6$, when this material is measured with the applied field being vertical to the $b$-axis \cite{lu2014superconductivity}. Until now, reported Ta-based superconductors with $\mu_0 H_{\rm c2}(0)$ value higher than the weak coupling Pauli limit are limited to: TaRh$_2$B$_2$, TaIr$_2$B$_2$, Ta$_2$Pd$_{0.92}$S$_6$, 4$H_{b}$-TaSe$_2$, Ta$_2$V$_{3.1}$Si$_{0.9}$, and the quasi-crystal superconductor Ta$_{1.6}$Te \cite{carnicom2018tarh2b2,gornicka2021nbir2b2,lu2014superconductivity,yan2023modulating,liu2023vanadium,terashima2024anomalous}. All these superconductors have highly anisotropic crystal structures. In contrary to this, the crystal structure of Ta$_4$Rh$_2$C$_{1-\delta}$ is cubic and centrosymmetric, which strongly reflects the unusual nature of the Pauli limit violation in this material. It should be noted that Ta$_4$Rh$_2$C$_{1-\delta}$ has the second highest critical temperature $T_{\rm c}$ and upper critical field $\mu_0 H_{\rm c2}(0)$ values among all reported $\eta$-carbide structure type superconductors, to date \cite{ma2021superconductivity, ma2021group, watanabe2023observation, ma2019superconductivity}.

\begin{table} [h!]
\caption{Summary of all the determined parameters of Ta$_4$Rh$_2$C$_{1-\delta}$ and the comparison with Nb$_4$Rh$_2$C$_{1-\delta}$. }
\begin{center}
\begin{tabular}{| c | c | c | c |}
\hline
Parameters & Units &Ta$_4$Rh$_2$C$_{1-\delta}$ & Nb$_4$Rh$_2$C$_{1-\delta}$ \\
\hline
$T_{\rm c,m}$ & K & 6.3  & 9.7 \\
$T_{\rm c,r}$ & K& 6.4 & 9.8 \\
$T_{\rm c,h}$ & K& 6.0 & 9.5 \\
RRR &  -& 1.22 & 1.16 \\
$\mu_0 H_{\rm c1}^{*}(0)$ & mT & 20& 13.6 \\
$\mu_0 H_{\rm c2}(0)$& T& 17.4 & 28.5 \\
$\beta$ & mJ mol$^{-1}$ K$^{-4}$& 0.7&  0.6 \\
$\gamma$ & mJ mol$^{-1}$ K$^{-2}$&21&40\\
$\lambda_{\rm ep}$&-& 0.71& 0.83\\	
 $\Theta_D$& K&268&283\\
$\xi_{GL}$ & \AA &  44& 34\\
$\lambda_{GL}$& \AA & 1743 & 2252 \\
$\kappa_{\rm GL}$ & -& 40 &  66.2 \\
$\Delta C/\gamma T_{\rm c}$ &-&  1.56& \ 1.64 \\
$D_{exp}$($E_{\rm F}$) & states eV$^{-1}$/f.u.& 5.23& 9.32\\
$D_{cal}$($E_{\rm F}$) & states eV$^{-1}$/f.u.& 5.45& 9.63\\
$\mu_0 H_{\rm c2}(0)$/$T_{c}$ & T/K &2.85&2.92 \\
\hline
\end{tabular}
\label{table:Table2}
\end{center}

{Note : $\mu_0 H_{\rm c1}^{*}(0)$ value for Ta$_4$Rh$_2$C$_{1-\delta}$ is corrected with demagnetization factor.} 
\end{table}

\subsection{Parameters in the superconducting state of Ta$_4$Rh$_2$C$_{1-\delta}$}

The obtained upper critical field $\mu_0 H_{\rm c2}(0) $ value, together with the lower critical field $\mu_0H_{\rm c1}^{*}(0)$ value can be used to calculate other relevant superconducting parameters for Ta$_4$Rh$_2$C$_{1-\delta}$. Here, the $\mu_0 H_{\rm c2}(0) $ value from the 50\%-criterion is 17.4 T, and it corresponds to a superconducting Ginzburg-Landau coherence length of $\xi_{\rm GL}$ $=$ 43.5 \AA \ according to the following equation:

\begin{equation}
\mu_0H_{\rm c2}(0) = \frac{\Phi_0}{2 \pi \ \xi_{\rm GL}^2}.
\label{eq:GL}
\end{equation}
where $\Phi_0 = h/(2e) \approx 2.0678 \times  10^{-15}$ Wb is the quantum flux. The superconducting penetration depth $\lambda_{\rm GL}$ can be estimated from the  values of $\xi_{\rm GL}$ and $\mu_0H_{c1}^{*}$ obtained above by using the relation:
 
\begin{equation}
\mu_0 H_{c1}^{*} = \frac{\Phi_0}{4 \pi \lambda_{\rm GL}^2} ln(\frac{\lambda_{\rm GL}}{\xi_{\rm GL}}).
\end{equation}

We obtained a value of $\lambda_{\rm GL} =$ 1743 \AA \ for Ta$_4$Rh$_2$C$_{1-\delta}$. The value of $\kappa_{\rm GL} = \lambda_{\rm GL}/\xi_{\rm GL}$ is calculated to be 40. These values demonstrate that Ta$_4$Rh$_2$C$_{1-\delta}$ is a type-II superconductor with a short superconducting coherence length $\xi_{\rm GL}$ and a large superconducting penetration depth $\lambda_{\rm GL}$.

In Table \ref{table:Table2}, we list all the parameters that we have obtained for Ta$_4$Rh$_2$C$_{1-\delta}$ and compare them with its isostructural superconductor Nb$_4$Rh$_2$C$_{1-\delta}$. We find the superconducting properties of Ta$_4$Rh$_2$C$_{1-\delta}$ are similar to those of Nb$_4$Rh$_2$C$_{1-\delta}$, especially the high upper critical field exceeding the weak coupling Pauli limit. Therefore, Ta$_4$Rh$_2$C$_{1-\delta}$ is both isostructural and isoelectronic to its sister compound Nb$_4$Rh$_2$C$_{1-\delta}$. Previously, the isostructural and isoelectronic Nb/Ta superconducting sister compounds pairs NbC - TaC \cite{shang2020superconductivity}, NbRh$_2$B$_2$ - TaRh$_2$B$_2$ \cite{carnicom2018tarh2b2}, and NbIr$_2$B$_2$ - TaIr$_2$B$_2$ \cite{gornicka2021nbir2b2} have been explored and compared. Here, Ta$_4$Rh$_2$C$_{1-\delta}$ and Nb$_4$Rh$_2$C$_{1-\delta}$ represent a new pair of isostructural and isoelectronic Nb/Ta superconducting compounds.

\begin{figure}
\centering
\includegraphics[width=0.6\linewidth]{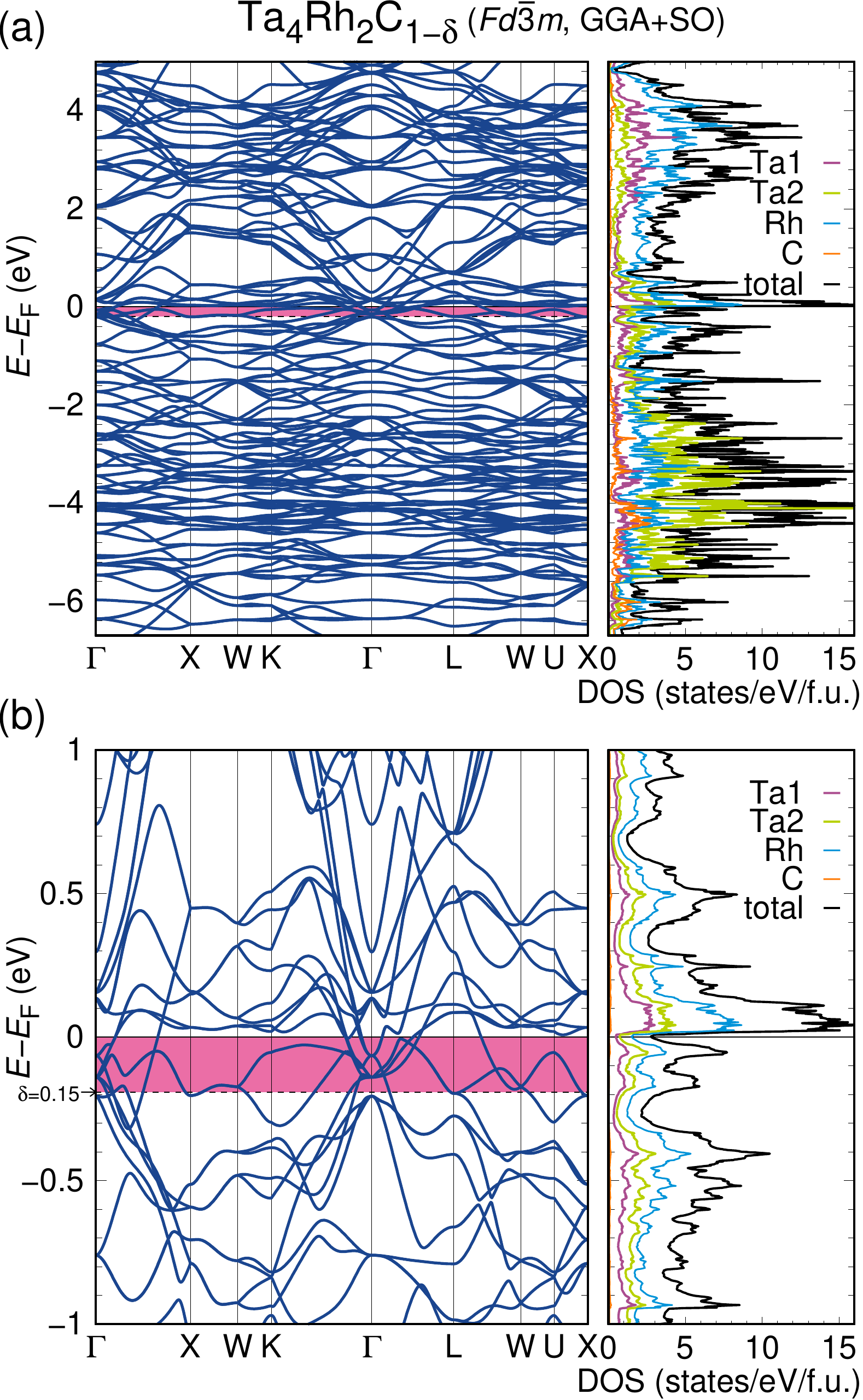}
\caption{Electronic structure of Ta$_4$Rh$_2$C$_{1-\delta}$ calculated with GGA+SO. (a) Overview and (b) detail of the band structure and density of states. Pink shading indicates the range $0 \le \delta \le 0.15$.}
\label{fig4}
\end{figure}

\subsection{Electronic structure of Ta$_4$Rh$_2$C$_{1-\delta}$} 

We performed density functional theory calculations for Ta$_4$Rh$_2$C$_{1-\delta}$  based on the structure determined in this work and given in Table~\ref{table:Table1}.
Figure~\ref{fig4} shows the band structure and density of states (DOS) of Ta$_4$Rh$_2$C$_{1-\delta}$. We perform the calculation for $\delta=0$, but we determine the amount of hole doping a carbon deficiency of $\delta=0.15$ would entail; it is marked by a pink region in Figure~\ref{fig4}. As Ta is chemically similar to Nb, it is not surprising that the electronic structure of Ta$_4$Rh$_2$C$_{1-\delta}$ does resemble the electronic structure of Nb$_4$Rh$_2$C$_{1-\delta}$ (see Ref. \cite{ma2021superconductivity}). From the calculated DOS shown in Figure~\ref{fig4} (b), the DOS at $E_{\rm F}$ with $\delta=0.15$ gives a theoretical value of 5.45 states eV$^{-1}$/f.u., which is comparable with the value of 5.23 states eV$^{-1}$/f.u. calculated from heat capacity measurement (see Table \ref{table:Table2}). Similarly, in the previous study on Nb$_4$Rh$_2$C$_{1-\delta}$, the DOS at $E_{\rm F}$ with $\delta=0.3$ provided a theoretical value of 9.63 states eV$^{-1}$/f.u., which is close to the experimentally derived value of 9.32 states eV$^{-1}$/f.u. from heat capacity measurement(see Ref. \cite{ma2021superconductivity}). 

We have also compared the effect of spin-orbit coupling on the band
structures of Nb$_4$Rh$_2$C and Ta$_4$Rh$_2$C, respectively. We find significantly
stronger splitting of bands in Ta$_4$Rh$_2$C, indicating stronger effects of
spin-orbit coupling due to the replacement of the $5d$ transition metal Ta for the $4d$ transition metal Nb (see supplementary information S-Figure 8) \cite{supp}. In the supporting information (S-Figure (9) and (10)), we present the calculated Fermi surface of Nb$_4$Rh$_2$C$_{1-\delta}$ with $\delta=0$, and $\delta=0.15$, respectively \cite{supp}.

\section{Conclusion}

In summary, we have successfully synthesized a new  $\eta$-carbide superconductor Ta$_4$Rh$_2$C$_{1-\delta}$ by arc-melting followed by the high-temperature annealing method. Our X-ray diffraction measurements show that Ta$_4$Rh$_2$C$_{1-\delta}$ crystallizes in the $\eta$-carbide structure type, and is isostructural to the known superconductor Nb$_4$Rh$_2$C$_{1-\delta}$. Our systematic temperature dependent magnetic susceptibility, resistivity, and specific heat capacity measurements show Ta$_4$Rh$_2$C$_{1-\delta}$ is a bulk superconductor with a critical temperature of $T_{\rm c}$ of 6.4 K, and a specific heat jump value $\Delta C/\gamma T_{\rm c}$ of 1.56, close to the weak-coupling BCS value of 1.43. 

It is indeed an extreme type-II superconductor with a $\kappa_{\rm GL} = \lambda_{\rm GL}/\xi_{\rm GL}$ to be 40. The upper critical field $\mu_0 H_{\rm c2}(0)$ of 17.4 T is exceeding the weak-coupling BCS Pauli paramagnetic limit of $\mu_0 H_{\rm Pauli} =$ 11.9 T. All these intriguing properties make Ta$_4$Rh$_2$C$_{1-\delta}$ an exotic superconductor similar to its sister compound Nb$_4$Rh$_2$C$_{1-\delta}$. In the future, the development of improved preparation techniques to obtain single phase or even single crystal samples of these $\eta$-carbide type superconductors for the development of superconducting wires, but also for an improved understanding of the underlying superconducting mechanism, is highly desired.

\begin{acknowledgments}
This work was supported by the Swiss National Science Foundation under Grant No. PCEFP2\_194183. Research performed at Gdansk Tech. was supported by the National Science Center (Poland), Project No. 2022/45/B/ST5/03916
\end{acknowledgments}

\bibliography{apssamp}% Produces the bibliography via BibTeX.

\end{document}